\documentclass[11pt]{article}
\usepackage{color,bm,epsfig,graphics,amssymb,amsmath,subeqnarray,setspace,graphicx,amsthm,epstopdf,subfigure,color,enumerate,pgfgantt,xcolor,wrapfig}
\usepackage[margin=1in]{geometry}
\usepackage{graphicx}
\usepackage{amssymb,amsmath}
\usepackage{bm}
\usepackage{authblk}

\begin{document}

\setcounter{page}{1}

\newcommand{\taup}{\boldsymbol{\sigma}^p}
\newcommand{\sig}{\boldsymbol{\sigma}}
\newcommand{\De}{\textrm{De}}     
\newcommand{\Wi}{\textrm{Wi}}     
\renewcommand{\Re}{\textrm{Re}}   
\newcommand{\dgamma}{\dot{\boldsymbol{\gamma}}}
\newcommand{\f}{\mathbf{f}}
\renewcommand{\u}{\mathbf{u}}
\newcommand{\Id}{\mathbf{I}}
\newcommand{\eps}{\varepsilon}
\renewcommand{\Re}{\textit{Re}}
\newcommand{\F}{\mathbf{F}}
\newcommand{\U}{\mathbf{U}}
\renewcommand{\d}{\partial}
\newcommand{\C}{\mathbf{C}}
\newcommand{\T}{\mathbf{T}}
\renewcommand{\v}{\mathbf{v}}
\newcommand{\Ftan}{\mathbf{F}_{\parallel}}
\newcommand{\Fnorm}{\mathbf{F}_{\perp}}
\newcommand{\tr}{\textrm{Tr}}
\newcommand{\D}{\mathbf{D}}
\renewcommand{\S}{\mathbf{S}}
\newcommand{\Omid}{\Omega_{\textrm{mid}}}
\newcommand{\Otip}{\Omega_{\textrm{tip}}}
\newcommand{\W}{W}
\newcommand{\E}{\mathcal{E}}
\newcommand{\ucop}{\mathcal{G}}
\newcommand{\x}{\mathbf{x}}
\newcommand{\Uinf}{\mathbf{U}_{\infty}}
\newcommand{\ein}{\tr(\C-\Id)}
\newcommand{\einbo}{\tr(\C_0-\Id)}
\newcommand{\Str}{\mathcal{S}}
\newcommand{\Ftip}{F^{\textrm{tip}}}
\newcommand{\mDt}[1]{\mathcal{D}_t{#1}}
\newcommand{\ytip}{y_{\textrm{tip}}}
\newcommand{\sign}{\boldsymbol{\sigma}^n}
%
\definecolor{review_color}{rgb}{0.0, 0.0, 0.0}  
\newcommand{\change}[1]{\textcolor{review_color}{#1}}


\title{Orientation dependent elastic stress concentration at tips of slender objects translating in viscoelastic fluids}
\author[b]{Chuanbin Li}
\author[a]{Becca Thomases}
\author[a]{Robert D. Guy}
\affil[a]{Department of Mathematics, University of California Davis, Davis, CA 95616}
\affil[b]{Department of Mathematics, The Pennsylvania State University, University Park, PA 16802}
%


\date{\today}

\maketitle


\begin{abstract}
Elastic stress concentration at tips of long slender objects moving in viscoelastic fluids has been observed in numerical simulations, but despite the prevalence of flagellated motion in complex fluids in many biological functions, the physics of stress accumulation near tips has not been analyzed.   Here we theoretically investigate elastic stress development at tips of slender objects by  computing the leading order viscoelastic correction to the equilibrium viscous flow around long cylinders, using the weak-coupling limit. In this limit nonlinearities in the fluid are retained allowing us to study the biologically relevant parameter regime of high Weissenberg number. 
We calculate a stretch rate from the viscous flow around cylinders to predict when large elastic stress develops at tips, find thresholds for large stress development depending on orientation, and calculate greater stress accumulation near tips of cylinders oriented parallel to motion over perpendicular.


 \end{abstract}
 



\section{Introduction}\label{Sec:intro}


The interaction of slender objects such as cilia and flagella with surrounding viscoelastic fluid environments occurs in many important biological functions such as sperm swimming in mucus during fertilization and mucus clearance in the lungs.  
There has been much work devoted to understanding the effect of fluid elasticity in such systems including biological and physical experiments \cite{shen2011undulatory,liu2011force,espinosa2013fluid,qin2015flagellar}, asymptotic analysis for infinite-length swimmers  \cite{chaudhury_1979,fulford1998swimming,fu2007theory,fu2008beating,LAUGA:2007,fu2009swimming,dasgupta2013speed,riley2014enhanced,elfring2016effect}, and numerical simulations of finite-length swimmers \cite{teran2010viscoelastic,spagnolie2013locomotion,thomases2014mechanisms,salazar2016numerical,thomases2017role,li2017flagellar}. 
While flows around slender finite-length objects are essential to our understanding of the physics of micro-organism locomotion, our understanding of these flows in viscoelastic fluids is limited. Previous experimental and theoretical results have focused largely on sedimentation of slender particles in the 
limit of vanishing relaxation time, i.e. the low Weissenberg number limit \cite{leal1975slow,kim1986motion,chiba1986motion,joseph1996flow,huang1998direct,galdi2000slow,dabade2015effects}.  

\begin{figure}
\begin{center}
\includegraphics[width=.45\linewidth]{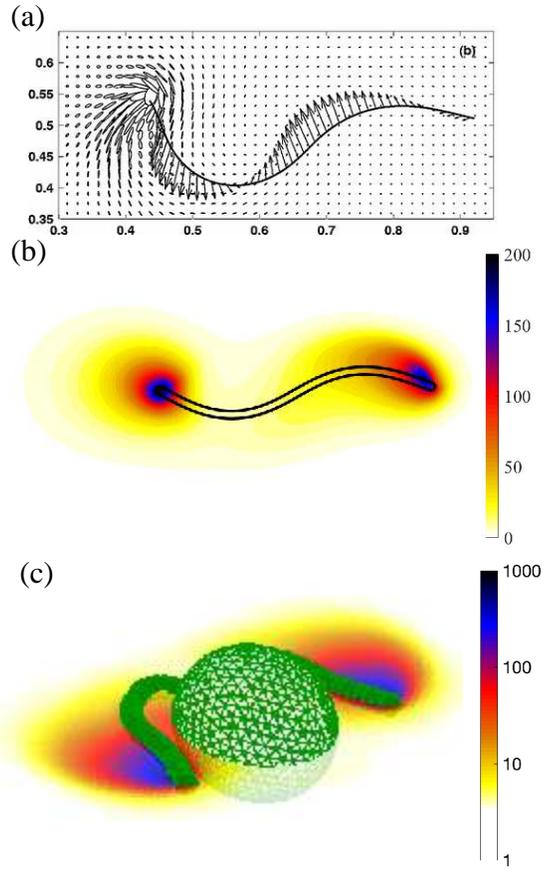}
\caption{(a) 2D flow around large amplitude, finite-length undulatory swimmer, ellipses show size and orientation of stress concentrated at tail  (reproduced from \cite{teran2010viscoelastic}). (b) 2D flow around bending sheet, color field shows strain energy density concentrated at tips. (c) 3D flow around swimming bi-flagellated cell, color field in center plane shows strain energy density.}
\label{fig:tip_stress}
\end{center}
\end{figure}

Numerical simulations of flagellated swimmers in viscoelastic fluids have shown the concentration of polymer elastic stress at the tips of slender objects  \cite{teran2010viscoelastic,thomases2014mechanisms,thomases2017role,li2017flagellar},  (see Fig.~\ref{fig:tip_stress})
 but why the stress concentrates so strongly at tips, and the effect of these stresses on micro-organism locomotion is not understood.  
  Unlike asymptotic theory \cite{fu2007theory,fu2008beating,LAUGA:2007,fu2009swimming,riley2014enhanced,elfring2016effect} these simulations involve large amplitude motions of finite length objects, and these large elastic stresses that arise have a substantially different effect on swimming motion than predicted by asymptotic analysis \cite{thomases2017role}.  
Experiments can measure kinematic changes \cite{shen2011undulatory,qin2015flagellar}, but not elastic stress, and thus the mechanisms of observed behavioral responses cannot be explained by experiments alone. 

It was observed in simulations \cite{li2017flagellar} that the concentrated tip stresses are stronger for a cylinder moving parallel to its axis compared to a cylinder moving perpendicular to its axis. 
This orientation dependence of elastic stress at tips is reversed from the orientation dependence of force on velocity in resistive force theory
and related viscous fluid theories \cite{hancock1953self,gray1955propulsion, keller1976slender,lighthill1976flagellar,johnson1979flagellar,johnson1980improved}
which form the basis of much of our intuition about micro-organism locomotion without inertia.
Classical viscous theories do not include tip effects, but 
previous results in viscoelastic fluids \cite{teran2010viscoelastic,thomases2014mechanisms,thomases2017role,li2017flagellar} suggest that 
the tip has a special role in the elastic stress development which has not been previously analyzed.

Previous work on the flow of viscoelastic fluids around slender objects has been done in the weakly nonlinear (or low Weissenberg number, $\Wi$) regime \cite{leal1975slow,kim1986motion,chiba1986motion,joseph1996flow,huang1998direct,galdi2000slow,dabade2015effects}, but the large stress concentration at tips of thin objects is a nonlinear effect and thus cannot be captured in a low $\Wi$ expansion. However, the highly nonlinear regime is challenging for numerical simulations \cite{owens2002computational}, and this has limited the ability to probe dynamics in this regime. 
\change{As another approach, one can consider the limit
of low polymer concentration, decoupling the stress and velocity.}
%
%
%
This method has been used to study stress localization for
high-$\Wi$ at extensional points and around objects \cite{rallison1988we,harlen1990high2, harlen1990high,renardy2006comment, thomases2007emergence}. 

The weak coupling expansion, a formal asymptotic approximation in the limit of low polymer concentration, retains viscoelastic nonlinearities at leading order \cite{moore2012weak}. This method has been successful in capturing high $\Wi$ effects for flow around a sphere in 3D \cite{moore2012weak}, and in the study of the rheology of dilute suspensions  in the low polymer concentration limit \cite{koch2016stress}.   
Similar stress localization in the wake of spheres has been observed experimentally \cite{acharya1976flow,arigo1998experimental} and theoretical predictions of shear thickening for strongly elastic dilute suspensions were in agreement with experimental observations \cite{scirocco2005shear}.
 
Here we use the weak coupling expansion to study the equilibrium flow around, and resultant force on, cylinders translating either perpendicular, or parallel to the direction of motion, in a 3D viscoelastic fluid.  Using this analysis we explain the origin of the tip stresses, we predict a critical Weissenberg number for the flow transition based on viscous flow data, and we show how the tip stress accumulation depends on cylinder orientation. 

   \begin{figure}
\begin{center}
\includegraphics[width=0.85\linewidth]{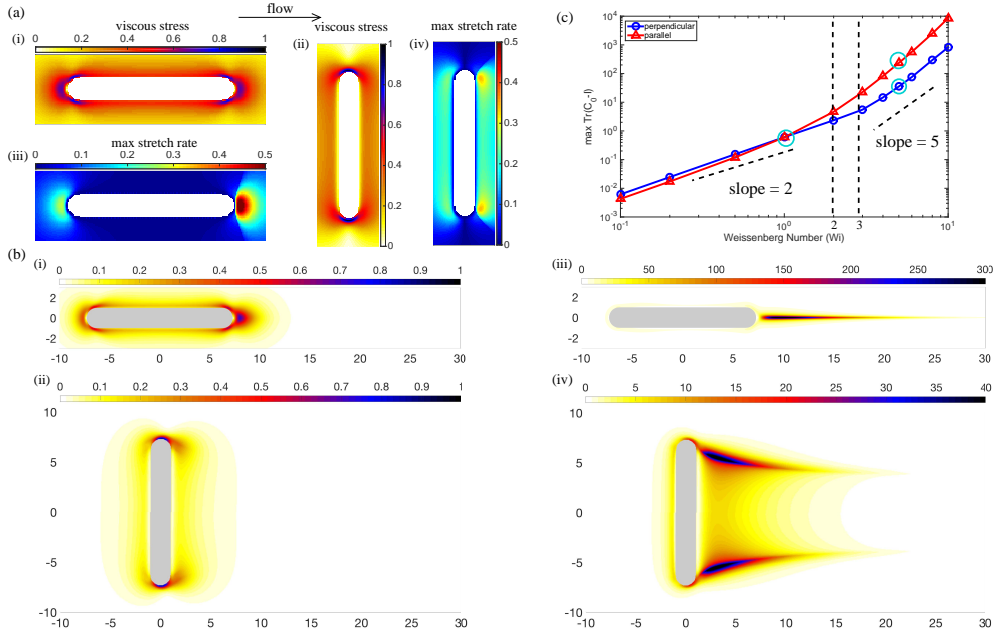}
\caption{(a) Norm of viscous stress in the center plane for cylinders oriented (i) parallel ($\max\|2\D^{\parallel}_0\|\approx 0.78$) and (ii) perpendicular ($\max\|2\D^{\perp}_0\|\approx 0.95$)  to flow; stretch rates of viscous flow for cylinders oriented (iii) parallel ($\max\nu^{\parallel}\approx 0.5$) and (iv) perpendicular ($\max\nu^{\perp}\approx 0.34$) to flow. Flow goes from left to right.  (b) $\einbo$ in the center plane for cylinders with (i)-(ii) $\Wi = 1$ and (iii)-(iv) $\Wi = 5$  (note the difference in scale). (c) Maximum of $\einbo$ as a function of $\Wi$ for the two orientations, in log scales. Dotted lines show the two critical Weissenberg numbers $\Wi\approx 2,$ and $\Wi\approx 3,$ cyan circles indicate $\Wi$ values pictured in (b).  }
\label{fig:stress}
\end{center}
\end{figure}

\section{Model Equations}

We examine the viscoelastic fluid flow around a stationary finite-length cylinder of radius $a$ with hemispherical caps driven
by a fixed flow at infinity, $\Uinf$.  We use the Oldroyd-B model of a viscoelastic fluid at zero Reynolds number, which is
attractive as a frame-invariant, nonlinear, continuum model of a viscoelastic fluid that can capture the dominant effects of fluid elasticity, e.g. storage
of history of deformation on a characteristic time-scale. 
The dimensionless system of equations is given by 
\begin{align}
&\Delta\u - \nabla p +\beta\nabla\cdot\C=0, \label{Stokes_main}\\
&\nabla\cdot\u=0,\label{incompressible_main}\\ 
 &\mDt{[\u]}\C=\Wi^{-1}\Id+\left(\nabla\u\C+\C\nabla\u^{T}\right)-\Wi^{-1}\C,\label{ucderiv_main}
\end{align} 
for $\u$  the fluid velocity,  $p$ the fluid pressure, and
$\C,$ the conformation tensor, a macroscopic average of the polymer orientation and stretching that is related to the polymer stress tensor by
 $\taup=\beta(\C-\Id)$. We use $\mDt{[\u]}$ to denote the material time derivative along the velocity field $\u.$
The parameters, $\beta,$ the non-dimensional polymer stiffness, and $\Wi,$ the Weissenberg number, 
or non-dimensional relaxation time, are defined by
\begin{equation}\label{betadef}\beta =\frac{Gr}{\mu U},\;\; \Wi =\frac{\lambda U}{r},\end{equation}
for $\mu$ the fluid viscosity,  $\lambda$ the fluid relaxation time, $G$ the polymer elastic modulus, and $U=|\Uinf|.$ 

The force on a stationary cylinder in a background flow is 
  proportional to the rate   at which energy is dissipated by the fluid. 
 To calculate the dissipation rate 
 we integrate the dot product of $(\u-\Uinf)$ and  Eq.~\eqref{Stokes_main} over the fluid domain,  $\Omega$ (exterior to the cylinder). After some manipulations 
and using the incompressibility constraint we obtain
  \begin{equation}\label{Fbal1}\Uinf\cdot\F = 2\int_{\Omega}D_{ij}D_{ij}\;dV+\beta\int_{\Omega}\frac{\partial u_i}{\partial x_j}C_{ij}\;dV,\end{equation} 
   where $D_{ij}=\frac{1}{2}\left(\frac{\d u_i}{\d x_j}+\frac{\d u_j}{\d x_i}\right)$ is the rate of strain tensor, $\F=\int_{\d\Omega}(\sign+\beta\C)\cdot\mathbf{n}\;dS$ is the force on the cylinder, and $\sign=2\D-p\Id$ is  the Newtonian stress tensor.   Thus for a constant velocity at infinity the force on the cylinder is proportional to
    the sum of the viscous dissipation rate and the rate at which energy is transferred to the polymers. 
   
 The polymer strain energy is  $\E =\int_{\Omega} \ein\;dV$ \cite{bird1977dynamics}, and an equation for the strain
 energy is obtained by taking the trace of Eq.~\eqref{ucderiv_main} and integrating over the fluid domain,
  \begin{equation}\label{energy_main}\frac{d}{dt}\E= 2\int_{\Omega}\frac{\d u_i}{\d x_j}C_{ij}\;dV-\Wi^{-1}\E.\end{equation}  
 Changes in the polymer energy come from transfer of energy between the fluid and the polymer and energy 
 lost to polymer relaxation. Therefore at steady state the rate of energy loss to the fluid is proportional to the polymer energy. 
By combining Eq.~\eqref{Fbal1} with Eq.~\eqref{energy_main} one finds that at steady state the force
on the cylinder is 
 \begin{equation}\label{Fbal2}\Uinf\cdot\F=2\int_{\Omega}D_{ij}D_{ij}\;dV+\frac{\beta}{2\Wi}\E.\end{equation} 
 Hence the strain energy $\E$ quantifies the force on the cylinder due to viscoelasticity.

\section{Weak-coupling expansion}
 Previous theoretical results on the polymeric contribution to a translating cylinder have used a second-order fluid expansion in the  weakly nonlinear regime \cite{leal1979motion,happel2012low, stone1996propulsion, dabade2015effects, elfring2015note}, where the nonlinearities associated with viscoelasticity are lost at leading order.
We are interested in the regime of large amplitude motions where large stress accumulates in the fluid, so we consider the \emph{weakly coupled}, or small $\beta,$ regime where the nonlinearities enter at leading order but the coupling between the polymer and fluid is higher order. 
The weak coupling 
  expansion was introduced for flow around a sphere in \cite{moore2012weak},
%
  %
   and is similar to analysis of viscoelastic fluids using fixed velocity fields in the high $\Wi$ regime \cite{renardy2000asymptotic,wapperom2005numerical}. Analysis of viscoelastic fluids with fixed velocity fields have predicted transitions in behavior for high $\Wi$ 
   at steady extensional points \cite{rallison1988we,harlen1990high2, harlen1990high,renardy2006comment, thomases2007emergence} and qualitatively similar transitions are also found in simulations where the velocity and the stress are fully coupled \cite{thomases2007emergence}.

 We expand the solutions in $\beta,$ $\u\sim\u_0+\beta\u_1,$  $p\sim p_0+\beta p_1,$ and $\C\sim\C_0+\beta\C_1.$   
At leading order Eqs.~\eqref{Stokes_main}--\eqref{incompressible_main} decouple from Eq.~\eqref{ucderiv_main}, and $\u_0$ is the
solution for the viscous flow around the cylinder. The conformation tensor satisfies  
  \begin{equation}\mDt{[\u_0]}\C_0=\Wi^{-1}\Id+\Str[\u_0]\C_0 -\Wi^{-1}\C_0,\label{ucderiv_b0_main}\end{equation}
where $\Str[\u_0]\C_0\equiv \left(\nabla\u_0\C_0+\C_0\nabla\u_0^{T}\right).$
On a given streamline Eq.~\ref{ucderiv_b0_main} is an ODE involving a source term, 
$\Wi^{-1}\Id$,  a stretching term, $\Str[\u_0] ,$ and a relaxation term, $\Wi^{-1}\C_0$.

\section{Tip stress development}

We prescribe a unit flow in the $x-$direction, $\U_{\infty}=\mathbf{e}_x,$ in the domain exterior to a cylinder that is oriented either parallel or perpendicular to the direction of flow, with no-slip boundary conditions on the cylinder walls.  The circular cylinder has length $4\pi,$ radius $a=1,$ and is capped at both ends with hemispheres.  We solve the Stokes equations for $\u_0$ using a boundary integral method based on a regularized Green's function from the method of regularized Stokeslets \cite{cortez2005method}. We generate streamlines of the Newtonian flow $\u_0$ and evolve Eq.~\eqref{ucderiv_b0_main} along those streamlines. See supplementary information for more details.

In Fig.~\eqref{fig:stress} (a) we plot the Frobenius norm (defined $\|\mathbf{A}\|\equiv\sqrt{A_{ij}A_{ij}}$) of the leading order viscous stress tensor $2\D_0$  in the center plane for cylinders oriented (i) parallel  and (ii) perpendicular to the flow. Note that the viscous stress near the middle of the cylinders is 2 or 3 times smaller than that at the tips.   In Fig.~\ref{fig:stress} (b) we show color fields of the leading order polymer strain energy density $\einbo,$
for two different Weissenberg numbers (i)-(ii) $\Wi=1$ and (iii)-(iv) $\Wi = 5.$   For $\Wi=1$ the elastic stress is concentrated at the tips like the viscous stress, and on the same scale as the viscous stress. For $\Wi = 5$ however, the elastic stress at the tips is more than 100 times larger than for $\Wi = 1,$ and concentrated in the wake. This nonlinear response 
has been seen before in analysis of flow around a circle in 2D \cite{hu1990numerical,mckinley1993wake,alves2001flow} and around a sphere in 3D \cite{moore2012weak}.  However, in Fig.~\ref{fig:stress} (b) (iii)-(iv) we also see that the stress in the wake of the cylinder that is oriented parallel to the direction of the flow is about 10 times larger than that for the cylinder oriented perpendicular to the direction of flow.  We examine the Newtonian flow that drives the stress growth to understand what sets the transition in $\Wi,$ and how the cylinder orientation impacts stress growth so  dramatically for large $\Wi.$  
  
At a fixed point in the flow, the real parts of the eigenvalues of the operator $\Str[\u_0],$ defined in Eq.~\eqref{ucderiv_b0_main}, set the growth (or decay) rates of $\C_0$ due to stretching (or compression).  \change{The solution to the eigenvalue problem $\Str[\u]\C=\nu\C$ is $\C=\v_i\v_j^T,$ $\nu_{ij}=\mu_i+\mu_j,$ where $\mu_i$ is an eigenvalue of $\nabla\u$ with corresponding eigenvector $\v_i.$}
We define the max stretch rate $\nu$ at a point as
\begin{equation}\label{stretchrate} \nu = 2\max(\textrm{Re}(\Lambda(\nabla\u_0))),\end{equation} where $\Lambda(A)$ is the set of eigenvalues of the matrix $A.$ 
In regions of the flow where $\nu-\Wi^{-1}>0,$ or $\nu\Wi>1,$ stretching outpaces relaxation, and while fluid particles remain in these stretching regions they experience unbounded stress growth.  

In Fig.~\ref{fig:stress} (a) we plot $\nu$ in the center plane for the cylinder oriented parallel (iii) and the cylinder oriented perpendicular (iv). The maximum stretch rate for both cylinders occurs in the wake of the cylinder, i.e. the max stretch rate contains information about flow directionality that is missing from Fig.~\eqref{fig:stress} (a) (i)-(ii). We see that the cylinder oriented parallel to motion has $\max(\nu^{\parallel})\approx 0.5,$ thus $\Wi^{\parallel}\approx 2$ is a threshold for stretching outpacing relaxation in regions of this flow. The maximum for the perpendicularly oriented cylinder is smaller, $\max(\nu^{\perp})\approx 0.34,$ corresponding to a threshold $\Wi^{\perp}\approx 3$ for large stress growth. 
For the perpendicularly oriented cylinder, the flow in the regions of high viscous stress near the tip is locally a shear flow, whereas, the local flow is extensional (which is known to lead to more rapid elastic stress growth \cite{larson1999structure}) near the tips of the cylinder oriented parallel.  The difference in flow type is reflected in the max stretch rate which is largest near the trailing tip of the cylinder oriented parallel to motion where the viscous stress is largest. For the perpendicularly oriented cylinder, the strongest extension is behind the cylinder where the viscous stresses are weaker. 

In Fig.~\ref{fig:stress} (c) we plot $\max\einbo$ for $\Wi\le10.$ For both orientations, the maximum of $\einbo$ scales like $\Wi^2$ below $\Wi\approx 2,$ and scales like $\Wi^5$ above $\Wi\approx 3$.
The cylinder oriented parallel to motion has a larger max stretch rate and thus
 it enters the regime of large stress growth for lower $\Wi$ than the perpendicularly oriented cylinder, leading to larger stress for a fixed $\Wi$ beyond the threshold $\Wi^{\parallel}\approx 2.$ 
Recall that the contribution to the force from the polymeric stress scales like $\frac{\beta}{2\Wi}\mathcal{E}$ and thus for low $\Wi$ there is a $\mathcal{O}(\Wi)$ contribution to the force whereas for high $\Wi$ the contribution is $\mathcal{O}(\Wi^4).$ Theoretical results have predicted similar scalings for related problems \cite{renardy2000asymptotic,wapperom2005numerical,moore2012weak}.

\section{Viscoelastic correction to force}

We expand the force on a cylinder to first order in $\beta$ as 
   \begin{equation}\label{Eq:Force}\F\sim\int_{\partial\Omega}\sign_0\cdot\mathbf{n}+\beta\left(\sign_1+\C_0\right)\cdot\mathbf{n}\;dS\equiv \F_0+\beta \F_1.\end{equation}    We avoid computing $\u_1,$ the first-order correction to the velocity, by using reciprocal relations \cite{leal1979motion,happel2012low, stone1996propulsion, dabade2015effects, elfring2015note}, as has been done before in many calculations of non-Newtonian corrections at low Reynolds number. 
 \change{ In addition, because the flow and force are parallel for these orientations, we obtain the magnitude of $\F_1$ as }
%
 %
  \begin{equation}\label{Eq:F1}F_1=\Wi^{-1}\int_{\Omega}\tr(\C_0-\Id)\;dV.\end{equation} 
     Details of our calculation are provided in supplementary information.
  Thus the viscoelastic correction to the force is proportional to the integral of the trace of the leading order polymer stress tensor over the fluid domain.  

\begin{figure}
\begin{center}
\includegraphics[width=.75\linewidth]{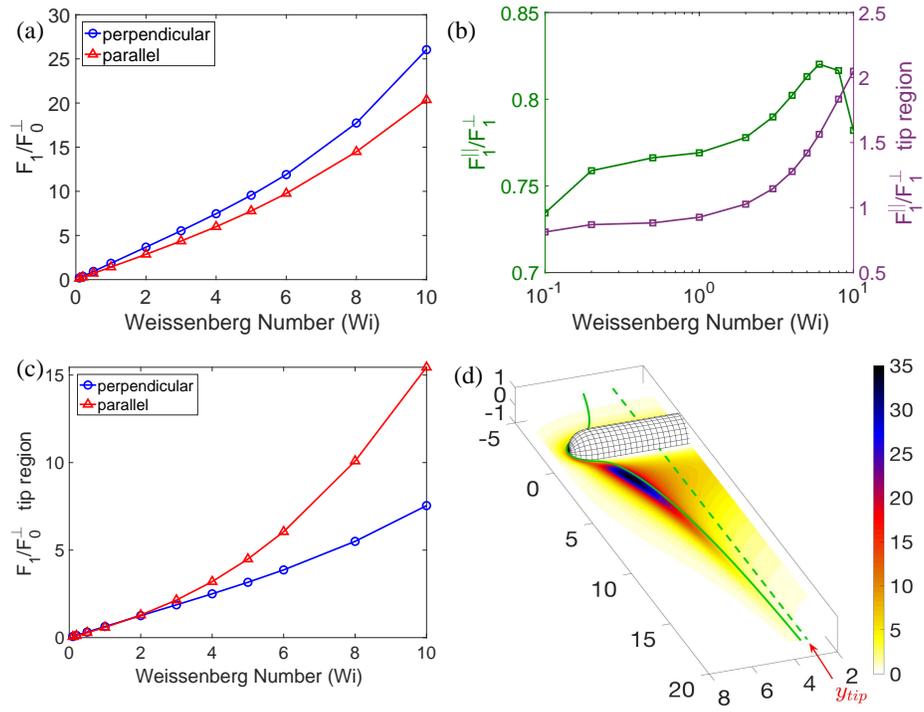}
\caption{(a) $F_1/F_0^{\perp}$ (perpendicular and parallel) (b)  parallel to perpendicular ratio of $F_1:$ whole domain (left axes), tip region (right axes) (c) $F_1$ restricted to tip  (perpendicular and parallel) (d) Diagram illustrating definition of $\ytip.$}
\label{fig:force}
\end{center}
\end{figure}

In Fig.~\ref{fig:force} (a) we plot the viscoelastic force correction, $F_1,$ normalized by $F_0^{\perp}=65$  (note $F_0^{\parallel}=48$)
for $\Wi\le10$ for each cylinder orientation. 
We see that in the expansion the $\mathcal{O}(\beta)$ force correction is up to 25 times the viscous force for large $\Wi.$ The perpendicular force correction is larger than the parallel force correction, however Fig.~\ref{fig:force} (b) shows  $F_1^{\parallel}/F^{\perp}_1$  (left hand axes) and beyond $\Wi\approx 2$ (the parallel stress growth threshold) $F_1^{\parallel}$ increases more than $F_1^{\perp},$ and this continues until about  $\Wi\approx 6$ where the ratio starts to decrease again. Since we are interested in the ``tip effect", we calculate the contribution to the
force from a single tip.

We define this tip force by restricting the integration domain in Eq.~\eqref{Eq:F1} to a subdomain 
exterior to the cylinder that contains only one tip. In Fig.~\ref{fig:force} (d) we show the tip of the perpendicular cylinder with the strain energy density for $\Wi=5.$ We consider a streamline that approaches very close to the tip in the center plane and we evolve the streamline until it levels off for large $x,$ and we define the value it approaches,  $\ytip=3.41,$ as shown in Fig.~\ref{fig:force} (d). With this we define 
\begin{equation}\label{ftipdef}\Ftip_1 =\Wi^{-1}\int_{\Omega\backslash\{y<\ytip\}}\einbo\;dV.\end{equation}
 In Fig.~\ref{fig:force} (c) we plot  $\Ftip_1/F_0^{\perp}$ for $\Wi\le10$ for each cylinder orientation, and the ratio of tip force corrections in Fig.~\ref{fig:force}(b) (right hand axes). Beyond the threshold $\Wi^{\parallel}\approx 2,$ the parallel force correction at the tip is larger than the perpendicular force correction, and the parallel force correction is double the perpendicular force correction from the tip at high $\Wi.$

\section{Discussion}\label{Sec:disc}
Using the viscous flow field around cylinders we predict a critical $\Wi$ beyond which a large stress ``tip effect" occurs, and we find that the critical $\Wi$ is orientation dependent. There are larger elastic stresses in the wake of cylinders oriented parallel to the direction of motion compared to cylinders oriented perpendicular to the flow. The flow type (shear or extensional) is orientation dependent and is reflected in the larger max stretch rate for the cylinder oriented parallel. The max stretch rate is defined from the eigenvalues of the operator $\Str$ in Eq. \eqref{ucderiv_b0_main}, and this operator appears in all differential models of viscoelasticity, including models which incorporate additional non-Newtonian effects such as shear-thinning. Hence we conjecture that the transitions we have identified are not specific to the Oldroyd-B model,
although quantitative values of stress accumulation beyond the transitions will depend on the model.

We explored other tip shapes and found that varying curvature at the tip did not effect the qualitative results; the max stretch rate was always largest near the tip, and greater for cylinders oriented parallel to the motion. The analysis given used the rod thickness to define the characteristic length scale and hence thinner rods will exhibit large stress growth at a lower relaxation time.  Although the tip effect is independent of the length, the relative
contribution to the total force from the tip depends on the length, 
and hence, quantifying the role of the tip effect on locomotion
requires more investigation. Nevertheless, based on past numerical
simulations of flagellated swimmers, it is clear that this tip effect
is significant.


In \cite{li2017flagellar} we observed elastic stress accumulation at flagellar tips in a simulation of a bi-flagellated alga cell swimming using experimentally measured kinematics.  
The stress accumulation was greater on the return stroke when the flagellar tips were oriented parallel to the direction of motion than when oriented perpendicular to the motion.  The steady-state analysis of the tip effect presented here
helps explain the physics behind these observations made in \cite{li2017flagellar}, but generally details of stroke kinematics, including time-dependence, will effect how stresses develop around flagellated swimmers.  We are able to make predictions about critical Weissenberg numbers for steady flows by looking at the max stretch rates,
but this tool could be useful for other gaits and even in experimental settings where flow fields are obtainable but location and concentration of stress are not measurable.

%


\vspace{.1in}
The authors thank David Stein for
interesting discussions and suggestions on this work. 
R.D.G. and B.T. were supported in part by NSF Grant No.
DMS-1664679.


%

\setcounter{page}{1}
\setcounter{section}{0}
\setcounter{figure}{0}
\setcounter{equation}{0}

%
%
%
%
%
%
%
%
%
%



\newcommand{\xb}{\mathbf{x}}
\newcommand{\Xb}{\mathbf{X}}
\newcommand{\ub}{\mathbf{u}}
%
\newcommand{\uv}{\mathbf{u}}

\pagestyle{plain}
\pagebreak
\begin{center}\LARGE
Supplemental Information for Orientation dependent elastic stress concentration at tips of slender objects translating in viscoelastic fluids\end{center}\normalsize
\vspace{.1in}
\begin{center}Chuanbin Li, Becca Thomases, and Robert D. Guy\end{center}
\vspace{.5in}

\renewcommand{\thepage}{S\arabic{page}}  
\renewcommand{\thesection}{S\arabic{section}}   
\renewcommand{\thetable}{S\arabic{table}}   
\renewcommand{\thefigure}{S\arabic{figure}}
\renewcommand{\theequation}{S\arabic{equation}}

\section{Numerical calculation of the flow and stress}

\subsection{Cylinder discretization}
The cylinders used in the computations are right circular cylinders of
length $L$ and radius $1$ with hemispherical caps on both ends. Thus
the tip-to-tip distance is $L+2$. Without loss of generality assume
that the long axis is oriented in the $x$-direction.
Let $N_{\theta}$ represent the number of points used to discretize the
circumference, and let $\Delta\theta = 2\pi/N_{\theta}$ be the spacing
between the points in the circumferential direction.
The axial direction is discretized with $N_{x}=L/\Delta\theta + 1 =
LN_{\theta}/(2\pi) + 1$ points so that the spacing along the axial
direction, $\Delta x$, is equal to the point spacing in the
circumferential direction. We choose $L$ to be an even multiple of
$\pi$ so that $N_{x}$ as defined above is an integer.
The caps are described in spherical coordinates. The azimuth angle is
discretized with the same spacing as the the circumferential angle,
i.e.\ $\Delta \phi = \Delta \theta$.

\subsection{Velocity}
We evaluate the velocity using a boundary integral with a regularized
kernel. The cylinder is stationary with velocity at $\infty$ given by
$\mathbf{U}_{\infty}=\mathbf{e}_{x}$. Using the boundary integral
formulation \cite{pozrikidis_book} the velocity can be evaluated by
\begin{equation}
  u_{i} (\xb) = \delta_{i,1} + \int_{\partial\Omega} G_{i,j}\left(\xb,\xb^{0}\right) F_{j}(\xb^{0}) \, dS\left(\xb^{0}\right),
  \label{BI_vel:eq}
\end{equation}
where the integral is over the cylinder surface. Because $\u=0$ on the
cylinder surface, the surface force density, $\mathbf{F}$, satisfies
the integral equation
\begin{equation}
  -\delta_{i,1} =\int_{\partial\Omega} G_{i,j}\left(\xb,\xb^{0}\right) F_{j}(\xb^{0}) \, dS\left(\xb^{0}\right).
  \label{BI_force:eq}
\end{equation}
Here $G_{i,j}\left(\xb,\xb^{0}\right)$ is the free-space Green's
function for Stokes equation representing the flow in the
$i$-direction at point $\xb$ generated by a unit point force oriented
in the $j$-direction at point $\xb^{0}$.

We replace the singular kernels by regularized kernels using the
Method of Regularized Stokelests  \cite{cortez2001method,cortez2005method}. The
discretized and regularized versions of equations \eqref{BI_vel:eq} and
\eqref{BI_force:eq} are, respectively
\begin{equation}
   u_{i} (\xb) = \delta_{i,1} + \frac{1}{8\pi\mu}\sum_{k} S_{i,j}^{\epsilon}\left(\xb , \xb_{k} \right) F_{j}(\xb_{k}) dA_{k},
  \label{RS_vel:eq}
\end{equation}
and
\begin{equation}
  -\delta_{i,1} =  \frac{1}{8\pi\mu}\sum_{k} S_{i,j}^{\epsilon}\left(\xb_{n} , \xb_{k} \right) F_{j}(\xb_{k}) dA_{k},
  \label{RS_force:eq}
\end{equation}
where $\xb_{n}$ and $\xb_{k}$ are points on the discretized
cylinder. We use the form of the regularized Stokeslet from
\cite{cortez2005method}, which is
\begin{equation}
  S_{i,j}^{\epsilon}\left(\xb,\xb^{0}\right) =
     \delta_{i,j}\frac{r^{2}+2\epsilon^2}{\left(r^2 +\epsilon^2\right)^{3/2}}
     + \frac{(x_{i}-x_{i}^{0})(x_{j}-x^{0}_{j})}{\left(r^2 +\epsilon^2\right)^{3/2}},
\end{equation}
where $r=\sqrt{(x_i-x_{i}^{0})(x_i-x_{i}^{0})}$. We set $\epsilon =
1.2\Delta x$, where $\Delta x$ is the point spacing along along the
cylinder in the axial direction, and the viscosity is set to $1$ from
the nondimensionalization.

\begin{figure}
  \centering
  \includegraphics[width=0.5\textwidth]{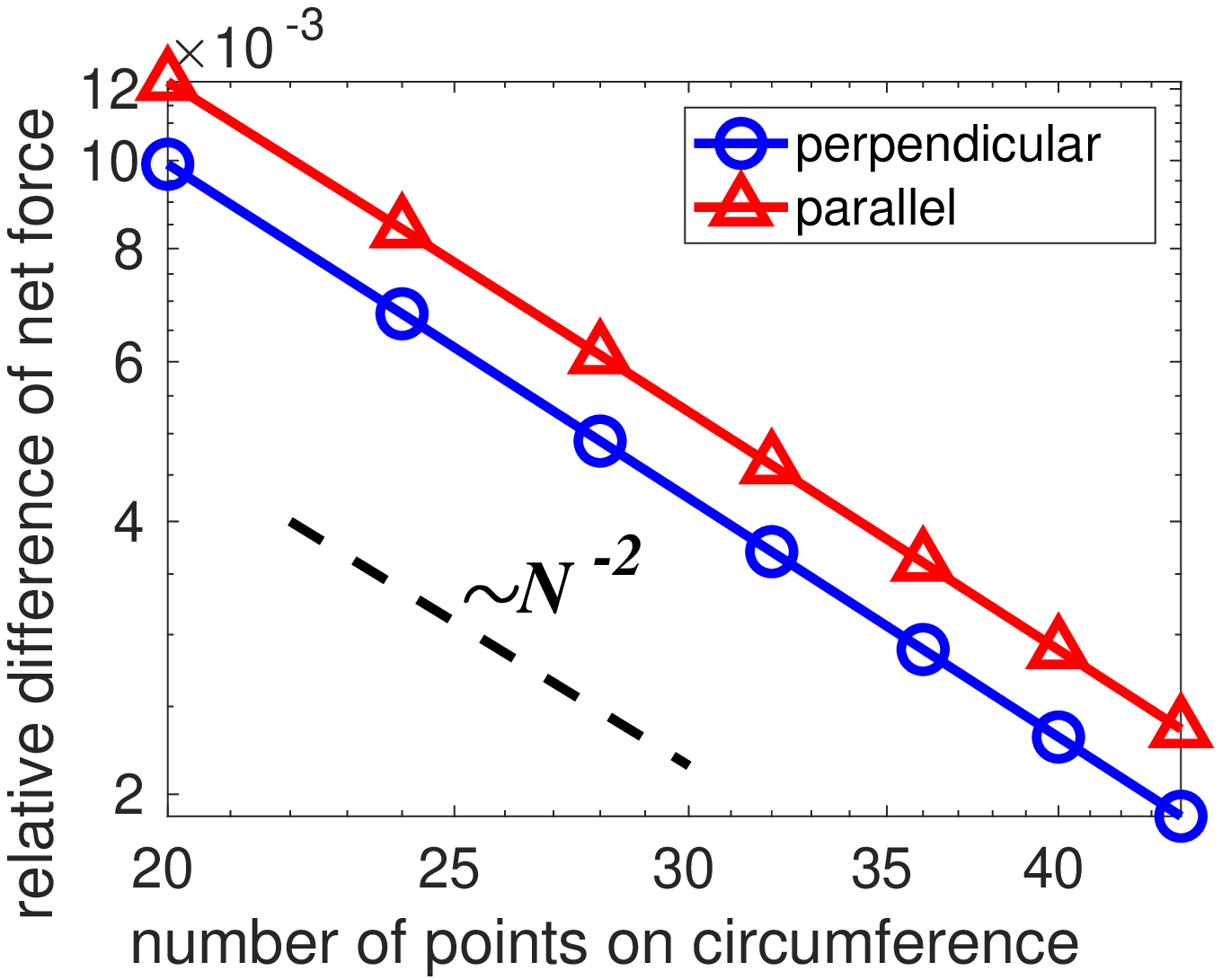}
  \caption{Refinement study for the magnitude of the viscous force on the
    cylinder as the number of points on the circumference increases.}
  \label{force_refine:fig}
\end{figure}

In Figure \ref{force_refine:fig} we show the results of a refinement
study for the magnitude of the viscous force as the number of points
on the circumference of the cylinder is refined. The plot shows that the
method converges at second-order in the point spacing along the
cylinder. The computations in the paper were performed with at least
24 points on the cylinder circumference. The error in the force is
estimated to be on the order of 1\% at this resolution using the
relative difference of the force between successive meshes as an
estimate of the error.

\subsection{Conformation tensor}
We identify streamlines by integrating the velocity field with a
relative error tolerance of $10^{-4}$. Along each streamline we
compute the velocity gradient using a second-order centered
finite-difference with points spaced $10^{-6}$ in each direction off
the streamline. In the weak coupling limit, the polymer stress
decouples from the flow, and so the conformation tensor is computed
from a known velocity gradient at each order in $\beta$. On a given streamline,
$\Xb(t)$, the the equation for the conformation tensor reduces to an
ODE, which at leading order is
\begin{equation}
  \frac{d \C_{0}}{dt} =\Wi^{-1}\Id +
    \bigl(\nabla\u_0(\Xb(t))\C_0
          +\C_0\nabla\u_0^{T}(\Xb(t))\bigr)
          -\Wi^{-1}\C_0.
\end{equation}
We use cubic splines to represent the velocity gradients along the
streamline, and we integrate this ODE on each streamline with a
relative tolerance of $10^{-4}$. The conformation tensor is
initialized to the identity tensor at the upstream boundary of the
computational domain.

\subsection{Integral of the trace of the stress}
We compute the solution in a $[-50, 50]^3$ box, but for each
orientation, we exploit the appropriate symmetry. The velocity
gradient decays like $r^{-2}$, and the viscous dissipation rate
density decays like $r^{-4}$. The viscous force on the cylinder
is the integral of the dissipation rate over the exterior of the
cylinder. The relative truncation error of using a domain of size $R$
is $\sim R^{-1}$, and so for a box of size $50$ we estimate the
truncation errors are around 2\%.

For the parallel orientation, the solution is axi-symmetric,
and so we compute the streamlines and stresses in the half plane
$y=0$, $z\geq 0$. The starting points for the streamlines are a set of
points along the line segment $x=0$, $y=0$ and $z\in(1,50]$. From
  these starting points, the streamlines are generated by integrating
  forward and backward in time until they reach the boundary of the
  computational box.

The starting positions of the streamlines are spaced more closely near
the cylinder surface than far away.  The velocity gradients decay
proportional to $r^{-2}$ far from the cylinder, and so few points are
needed away from the cylinder.  Specifically, we use the nonlinear
transformation
  \begin{equation}
    z = (1-b w)^{-1/b},
  \end{equation}
where $b\geq -1$ is a parameter that affects how clustered the points
are near the cylinder surface. Larger values of $b$ result in more
clustering of points near the cylinder.  This transformation maps the
interval $[0,w_{max}]$ to $[1,z_{max}]$, where
$w_{max}=(1-z_{max})^{-b}/b$. We choose a uniformly spaced
discretization in $w$, and then use the transformation to define the
starting $z$ values. The transformation is defined by the differential
equation
  \begin{gather}
    dz = z^{b+1} dw \\
    z(0) = 1.
  \end{gather}
Thus, the spacing grows as a power law in $z$ increases.  For
the computations in the manuscript, we use $b=0.75$ and $36$ discrete
points. With these parameters, the streamlines closest to the cylinder
are spaced about $0.036$ apart, which are much more closely spaced
than the points along the cylinder. There are 16 points within the
first unit of distance and just 5 points within the interval
$[10,50]$.

For the perpendicular orientation we use a similar approach for
increasing the streamline spacing away from the cylinder. The axis of
the cylinder is aligned in the $y$-direction. Because of the symmetry,
we compute the solution for $y,z\geq 0$. The starting points for
identifying streamlines are from the quarter plane $x=0$, $y,z\geq
0$. Streamlines are found from integrating forward and backward in
time until reaching the boundary of the computational box
($x=\pm50$). The starting points are located along curves of a fixed
distance from the cylinder surface. The spacing between these curves
is selected using the same discretization for starting values for the
parallel orientation described above. Along each curve,
points are chosen to be equally spaced with spacing approximately
equal to the spacing between the curves at this distance.

To compute integrals of the stress, we interpolate the stress to a
structured mesh. As with the streamline spacing the mesh is finer near
the cylinder and coarser away from it. For the parallel orientation case, we
begin with spacing in the $z$-direction on the interval $(1,50]$ using
  the same spacing as the starting points for the streamlines. We then
  add additional points in between any points which are spaced greater
  than $5$ units apart. For $z\in[0,1]$ we use equally spaced points
  with approximately the same spacing as the points near $z=1$.  The
  spacing in the $x$-direction is uniform with spacing approximately
  equal to the finest spacing in the $z$-direction near the cylinder.

For the perpendicular orientation, we use the same mesh for the
$z$-direction. For the $y$-direction (axial direction) we use uniform
spacing from $y=0$ (center) to $y=2\pi+1$ (tip) with the spacing equal
to the finest spacing in the $z$-direction. From the tip to the
computational boundary, we use the same stretched grid described
previously. In the region $y\in[2\pi+1,50]$ we use the same number of
points used in the in the interval $z\in[1,50]$. The spacing in the
$x$-direction is uniform with spacing approximately equal to the
finest spacing in the $z$-direction near the cylinder.

\begin{figure}
  \centering
  \includegraphics[width=0.5\textwidth]{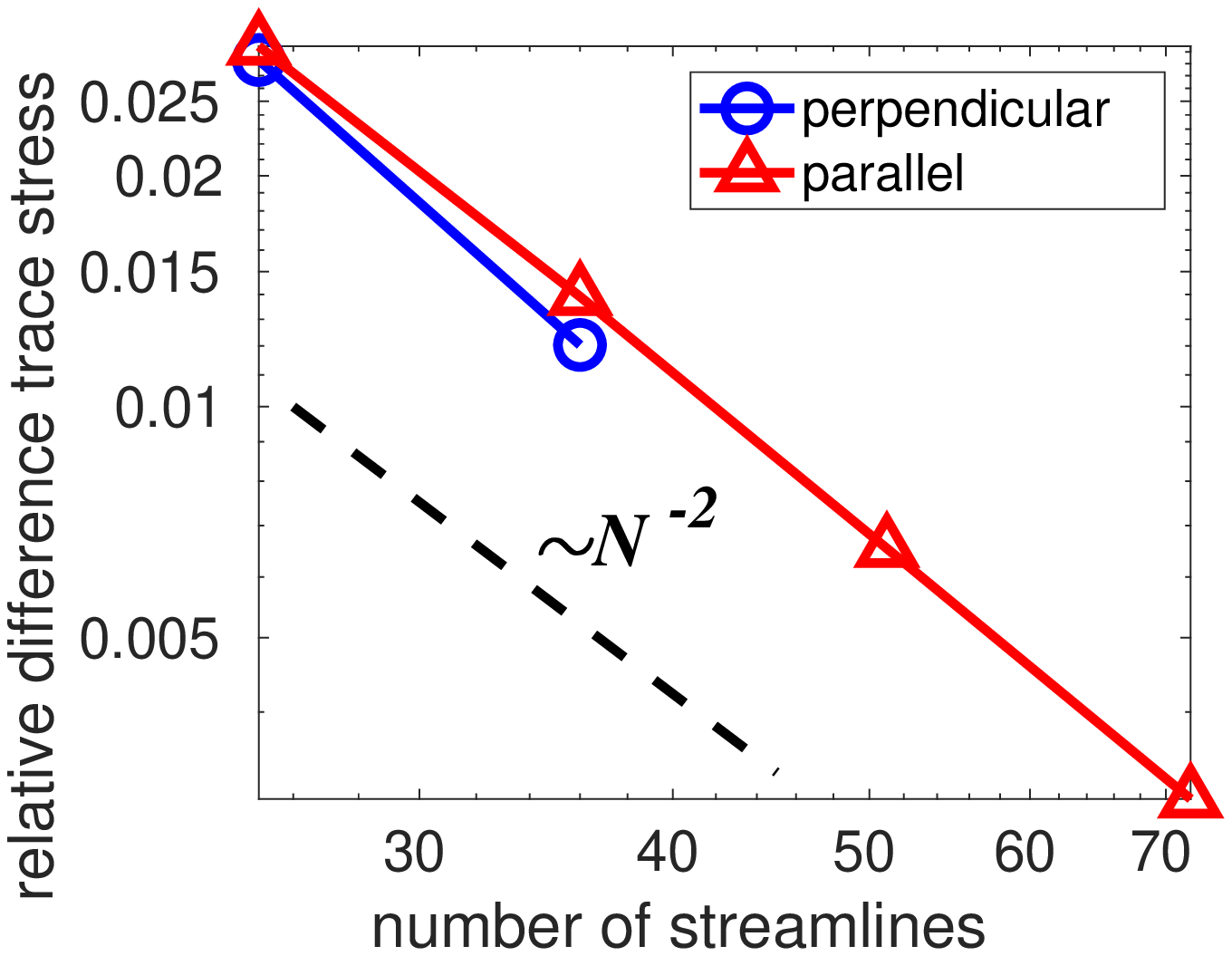}
  \caption{Refinement study for the integral of the trace of the stress
    for $Wi=1$ as the number of streamlines used along the $z$-axis
    increases. The number of points on the cylinder's circumference is
    24 for the perpendicular orientation and 28 for the parallel orientation.}
  \label{stress_refine_lam1:fig}
\end{figure}

After interpolating the stress to the mesh, the stress inside the
cylinder is set to zero, and the the stress is integrated in the cube
$[-50, 50]^3$ using trapezoidal rule. In Figure
\ref{stress_refine_lam1:fig} we show that this method converges at
second-order in the spacing between the streamlines. The refinement
study was performed by computing differences of the integral of the
trace of the stress between successive meshes. We used five meshes for
the parallel orientation, and three meshes for the perpendicular
orientation. The perpendicular case involves many more points and is
much more computationally expensive. In the paper we used 36
streamlines along the $z$-axis, and based on the refinement study, we
estimate that the errors are around 2\% at this resolution.

\section{Expression for force at first order}

\subsection{Reciprocal relation}

%
Let $\u$ be an incompressible velocity field and $\sig$ be
the associated Newtonian stress tensor. We consider two pairs of velocity and stress: $(\u,\sig)$ and $(\u',\sig')$.
\begin{align}
  u_{i}\frac{\partial \sigma'_{ij}}{\partial x_{j}}   & =
  \frac{\partial}{\partial x_{j}}\left( u_{i}\sigma'_{ij} \right)
  - \frac{\partial u_{i}}{\partial x_{j}}\sigma'_{ij} \\
  & =  \frac{\partial}{\partial x_{j}}\left( u_{i}\sigma'_{ij} \right)
  - \frac{\partial u_{i}}{\partial x_{j}}\left(-p'\delta_{ij} + \mu\left(\frac{\partial u'_{i}}{\partial x_{j}} + \frac{\partial u'_{j}}{\partial x_{i}}\right)\right) \\
  & = \frac{\partial}{\partial x_{j}}\left( u_{i}\sigma'_{ij} \right)
  - \mu \frac{\partial u_{i}}{\partial x_{j}}\left(\frac{\partial u'_{i}}{\partial x_{j}} + \frac{\partial u'_{j}}{\partial x_{i}}\right)
\end{align}
Now reversing the primes and subtracting we get
\begin{equation}
    u_{i}\frac{\partial \sigma'_{ij}}{\partial x_{j}}
    - u'_{i}\frac{\partial \sigma_{ij}}{\partial x_{j}} =
    \frac{\partial}{\partial x_{j}}\left(
    u_{i}\sigma'_{ij}- u'_{i}\sigma_{ij} \right).
    \label{reciprocal:eq}
\end{equation}
This identity will be the starting point for the manipulations below.


\subsection{Weak coupling expansion}

The dimensionless system of equations is
\begin{gather}
\Delta\u - \nabla p +\beta\nabla\cdot\C=0, \label{Stokes}\\
\nabla\cdot\u=0\label{incompressible}\\ 
 \frac{D\C}{Dt}=\Wi^{-1}\Id+\left(\nabla\u\C+\C\nabla\u^{T}\right)-\Wi^{-1}\C,\label{ucderiv}
\end{gather}
for $\u$ the fluid velocity, $p$ the fluid pressure, and $\C,$ the
conformation tensor, a macroscopic average of the polymer orientation
and stretching that is related to the polymer stress tensor by
$\taup=\beta(\C-\Id)$. The boundary conditions are $\u=0$ on the
cylinder surface, and $\u=\bm{e}_{x}$ at infinity. the net force on the
cylinder is 
\begin{equation}
  \bm{F} = \int_{\partial\Omega} (\sig^{n}+\beta\C)\cdot\bm{n}\,dS,
\end{equation}
where $\sig^{n}$ is the Newtonian stress.

Expanding the solution in powers of $\beta,$
Eqs. \eqref{Stokes}--\eqref{ucderiv} at leading order are
\begin{gather}
\Delta\u_0 - \nabla p_0 =0, \label{Stokes_b0}\\
\nabla\cdot\u_0=0, \label{incompressible_b0}\\ 
\frac{D\C_0}{Dt}=\Wi^{-1}\Id+ \left(\nabla\u_0\C_0+\C_0\nabla\u_0^{T}\right)
                 -\Wi^{-1}\C_0.\label{ucderiv_b0}.
\end{gather}
At first order the velocity and pressure satisfy
\begin{gather}
  \Delta\u_1 - \nabla p_1 =  -\nabla\cdot\C_{0}. \label{Stokes_b1}\\
  \nabla\cdot\u_1=0. 
\end{gather}
The force on the cylinder has the expansion
\begin{equation}
  \bm{F} \sim \int_{\partial\Omega} \sig_{0}^{n}\cdot\bm{n} + \beta(\sig_{1}^{n}+\C_{0})\cdot\bm{n} + \cdots dS
  = \bm{F}_{0} + \beta\bm{F}_{1} + \cdots .
\end{equation}
From this expression, it appears the velocity and pressure are needed
at first order to get the force at first order. However, using the
reciprocal relation the first order force can be obtained from the
leading order solution.

To use the reciprocal relation from Eq.\ \eqref{reciprocal:eq}, we
make the choice
\begin{gather}
  \u  = \u_{0}-\bm{e}_{x} \\
  \u' = \u_{1}
\end{gather}
and from Eqns.\ \eqref{Stokes_b0} and \eqref{Stokes_b1}, the respective stresses
satisfy 
\begin{gather}
  \nabla\cdot\sig = 0 \\
  \nabla\cdot\sig' = -\nabla\cdot\C_{0}.
\end{gather}
Plugging these into the reciprocal relation, Eq.\ \eqref{reciprocal:eq},
gives
\begin{equation}
  -\left(\uv_{0}-\bm{e}_{x}\right) \cdot \nabla\cdot\C_{0} =
  \nabla\cdot\bigl( \left(\uv_{0}-\bm{e}_{x}\right) \cdot \sig_{1}^{n} - \uv_{1}\cdot\sig_{0}^{n} \bigr)
\end{equation}
Because  $\bm{e}_{x}$ is constant, the above equation can be rearranged to
\begin{equation}
  -\u_{0}\cdot \nabla\cdot\C_{0} =
  \nabla\cdot\left( \u_{0}\cdot \sig_{1}^{n} -\bm{e}_{x}\cdot\left(\sig_{1}^{n}+\C_{0}\right)
   - \u_{1}\cdot\sig_{0}^{n} \right).
\end{equation}
Now integrate the above expression, apply the divergence theorem, use that 
$\u_{0}=\u_{1}=0$ on the surface and that the stresses are zero at infinity to get
\begin{equation}
  \int_{\Omega}-\u_{0}\cdot \nabla\cdot\C_{0} \, dV =
  \int_{\partial\Omega} \bm{e}_{x}\cdot\left(\sig_{1}^{n}+\C_{0}\right)\cdot\bm{n}\, dS
  = \bm{e}_{x}\cdot \bm{F}_{1}.
\end{equation}
Finally, integrate the left side by parts and use that $\uv_{0}=0$ on
the surface to get
\begin{equation}
  \bm{e}_{x}\cdot \bm{F}_{1} = \int_{\Omega} \nabla\uv_{0} :  \C_{0} \, dV.
  \label{F1_gradu_ddot_C:eq}
\end{equation}
This gives an expression for the force at first order in terms of
leading order quantities. We can manipulate this further using the
equation for the conformation tensor. The polymer strain energy is $\E
=\int_{\Omega} \ein\;dV$. By taking the trace of
Eqn.\ \eqref{ucderiv_b0}, the equation for the strain energy at
leading order is
\begin{equation}
  \label{energy}
  \frac{d}{dt}\E_{0}= 2\int_{\Omega}\frac{\d (u_{0})_i}{\d x_j}(C_{0})_{ij}\;dV
    -\Wi^{-1}\E_{0}.
\end{equation}  
Taking this equation at steady state, and combining it with
Eqn.\ \eqref{F1_gradu_ddot_C:eq} gives the expression for the force in
terms of the strain energy as
\begin{equation}
  \bm{e}_{x}\cdot \bm{F}_{1} =\frac{1}{2\Wi} \int_{\Omega}\einbo\; dV.
\end{equation}

%
%
%



\end{document}